\documentclass[superscriptaddress,twocolumn,pre]{revtex4-1}
\usepackage{amssymb,amsmath,amsthm,wrapfig,esint,afterpage,marginnote,afterpage,lipsum,color,tabularx}
\usepackage{times}
\usepackage{bm}
\usepackage{lineno}
\usepackage{siunitx}
\usepackage{graphicx}
\usepackage{lipsum}
\usepackage{color,sidecap,fancyhdr}
\usepackage[dvipsnames]{xcolor}
\usepackage{mathpazo}
\usepackage{textcomp}
\usepackage{nicefrac}
\usepackage{xfrac}
\usepackage[sort&compress]{natbib}
\usepackage{dcolumn,bm}
\usepackage{hyperref}
\usepackage{cleveref}
\usepackage{mathtools}
\usepackage{epsfig}
\usepackage{epstopdf}
\usepackage[authormarkuptext=name]{changes}
\usepackage{changes}
\definechangesauthor[color=blue, name={AY}]{AY}

\newcommand{\bra}[1]{\left(#1\right)}
\newcommand{\Bra}[1]{\left[#1\right]}
\newcommand{\BRA}[1]{\left\{#1\right\}}

\begin{document}

\title{Excitable solitons: Annihilation, crossover, and nucleation of pulses in mass-conserving activator-inhibitor media}

\author{Arik Yochelis}\email{yochelis@bgu.ac.il}
\affiliation{Department of Solar Energy and Environmental Physics, Blaustein Institutes for Desert Research (BIDR), Ben-Gurion University of the Negev, Sede Boqer Campus, Midreshet Ben-Gurion 8499000, Israel}
\affiliation{Department of Physics, Ben-Gurion University of the Negev, Be'er Sheva 8410501, Israel}

\author{Carsten Beta}
\affiliation{Institute of Physics and Astronomy, University of Potsdam, 14476 Potsdam, Germany}

\author{Nir S. Gov}
\affiliation{Department of Chemical and Biological Physics, Weizmann Institute of Science, Rehovot 76100, Israel}

\received{\today}

\begin{abstract}
	Excitable pulses are among the most widespread dynamical patterns that occur in many different systems, ranging from biological cells to chemical reactions and ecological populations. Traditionally, the mutual annihilation of two colliding pulses is regarded as their prototypical signature. Here we show that colliding excitable pulses may exhibit soliton-like crossover and pulse nucleation if the system obeys a mass conservation constraint. In contrast to previous observations in systems without mass conservation, these alternative collision scenarios are robustly observed over a wide range of parameters.
	{We demonstrate our findings using a model of intracellular actin waves since, on time scales of wave propagations over the cell scale, cells obey conservation of actin monomers.} The results provide a key concept to understand the ubiquitous occurrence of actin waves in cells, suggesting why they are so common, and why their dynamics is robust and long-lived.
\end{abstract}


\maketitle

\section{Introduction}

The study of propagating solitary pulses is a cross-disciplinary field of research with important applications in biological, chemical, and physical systems~\cite{akhmediev2008dissipative,purwins2010dissipative}.
Solitary waves are commonly distinguished by their collision properties~\cite{scott1973soliton,scott1975electrophysics}: \emph{solitons} if after collision of two pulses, two pulses emerge (particle-like identity) and \emph{dissipative solitons} or \emph{excitable pulses} if they are annihilated.
While solitons are often discussed in the context of conservative media, excitable pulses typically arise in dissipative systems that contain auto-catalytic or enzymatic terms~\cite{meron1992pattern,Cross1993}.
Owing to their universal properties, they emerge over a wide range of scales, e.g., in surface reactions~\cite{rotermund1991solitons,von1998subsurface}, gas discharge plasmas~\cite{bode1995pattern}, intracellular actin dynamics~\cite{Allard2013,deneke2018chemical}, cardiac rhythms~\cite{karma2013physics}, and neuroscience~\cite{izhikevich2007dynamical}.

Annihilation of excitable pulses after a collision is well understood and recognized as paramount for electrophysiological function, as it would be impossible to maintain directionality and rhythmic behavior under the reflection of action potentials~\cite{alonso2016nonlinear}.
However, in several experimental~\cite{rotermund1991solitons,von1998subsurface,santiago1997dissolution,willebrand1992periodic,shrivastava2018collision} and theoretical cases~\cite{tuckwell1979solitons,kosek1995collision,mornev1996soliton,aslanidi1999soliton,nishiura2005scattering,tsyganov2007waves,lautrup2011stability,young2019interactions}, it was shown that also soliton-like behavior can be observed in dissipative reaction--diffusion (RD) media --- a finding that is typically restricted to a narrow range in parameter space and, to date, is considered as an exotic exception to the prototypical annihilation of excitable pulses.

Here we show that soliton-like behavior can robustly emerge in excitable RD media if they obey a mass-conservation constraint.
In contrast to previous cases without mass conservation, no fine-tuning of parameters is needed to observe crossover and pulse nucleation upon collision.
In particular, neither proximity to an oscillatory onset~\cite{argentina2000head} nor non-local interactions~\cite{krischer1994bifurcation,mimura1998collision,coombes2007exotic} nor cross--diffusion~\cite{tsyganov2003quasisoliton} are required.
To underline this paradigmatic shift in our understanding of excitable media, we term pulses in this regime as {\it excitable solitons}.

Our results are particularly important to understand dynamical patterns in biological cells, where mass conservation is {often} a dominant feature. {While cells are open systems that grow and divide, the total copy number of proteins such as actin monomers, vary over time-scales that are long compared to the time-scale it takes actin waves to propagate over the cell scale.} A prominent example are intracellular actin-membrane waves~\cite{Gerisch2004,Bretschneider2009,Gerhardt2014}
that are associated with fundamental cellular functions and appear in many cell types~\cite{beta_intracellular_2017,inagaki2017actin}, including {\it Dictyostelium} cells~\cite{vicker_pseudopodium_1997,Gerisch2004,Gerhardt2014}, neutrophils~\cite{weiner_actin-based_2007}, and fish keratocytes~\cite{barnhart_adhesion-dependent_2011}.
We, therefore, demonstrate our findings using the generalized version of a recently developed RD model with mass conservation that successfully describes the dynamics of wave-like actin polymerization in circular dorsal ruffles~\cite{bernitt2017fronts}.
We find rich dynamics of pulses upon collision, exhibiting not only the common regime of \textit{annihilation}, but also soliton-like \textit{crossover} and \textit{pulse nucleation} over a wide range of parameters.

\section{Mass-conserved RD model}

We start with an RD case model that was formulated to study front dynamics of circular dorsal ruffles (CDR)~\cite{bernitt2017fronts}, which are waves of actin polymerization that propagate on the dorsal side of the cell membrane. We reduce this model to a simpler version that includes filamentous actin and an inhibitor of actin polymerization. The three species in this minimal version are: ({\it{i}}) Polymerized actin filaments (F-actin) that are organized in a network (dendritic-like) morphology, $N(x,t)$, ({\it{ii}}) Actin monomers (G-actin) $S(x,t)$, and ({\it{iii}}) an actin polymerization inhibitor, $I(x,t)$. In accordance with the CDR model, we employ actin mass-conservation~\cite{bernitt2017fronts}:
\[
\Omega^{-1}\int_\Omega \BRA{N(x)+S(x)} {\mathrm d}x=A,
\]
where $x\in \Omega$ is the spatial domain size and $A$ is constant. In comparison with the CDR model, we have excluded from the current model the additional reservoirs of polymerized actin in the cortex and in stress fibers. Adding them complicates the analysis and does not qualitatively change the nature of the solitary pulses, which are the focus of this study.

The continuum model in its dimensionless form reads~\cite{bernitt2017fronts}:
\begin{subequations}\label{eq:RD}
	\begin{align}
		\frac{\partial N}{\partial t} &= {\frac{N^2 S}{1+I}} -{N} +  {{D}_\text{N} \frac{\partial^2 N}{\partial x^2}}, \label{Factin}\\
		\frac{\partial S}{\partial t} &= {- \frac{N^2 S}{1+I} + N} + \frac{\partial^2 S}{\partial x^2}, \label{Gactin}\\
		\frac{\partial I}{\partial t} &= {k_\text{N} N} - {k_\text{I} I} + {D_\text{I} \frac{\partial^2 I}{\partial x^2}}, \label{Inhibitor}
	\end{align}
\end{subequations}
Eqs.~\ref{Factin},\ref{Gactin} describe the auto-catalytic polymerization process, converting monomers to filaments, which is inhibited by the presence of $I$, and with a constant rate of depolymerization. Eq.~\ref{Inhibitor} describes the recruitment of the inhibitor to the filamentous actin. The hierarchy of diffusion coefficients, along the membrane, is such that the monomers diffuse the fastest, while the effective diffusion of polymerized actin is slower and mostly occurs by the polymerization activity. The inhibitor diffuses the slowest as it is adsorbed to the membrane~\cite{bernitt2017fronts}: $D_\text{I}\ll D_\text{N} <1$. In fact, $D_\text{I}$ is not essential for what follows, but we keep it as it makes the comparison to the FitzHugh-Nagumo (FHN) model~\cite{argentina2000head} transparent. In addition, we chose $k_\text{I}<k_\text{N}$~\cite{bernitt2017fronts}, but this is not essential. We employ Neumann (no-flux) boundary conditions (BC), while similar results (not surprisingly) are obtained with periodic BC.

\section{Linear stability analysis of uniform solutions}

\begin{figure*}[tp]
	(a)\includegraphics[width=0.44\textwidth]{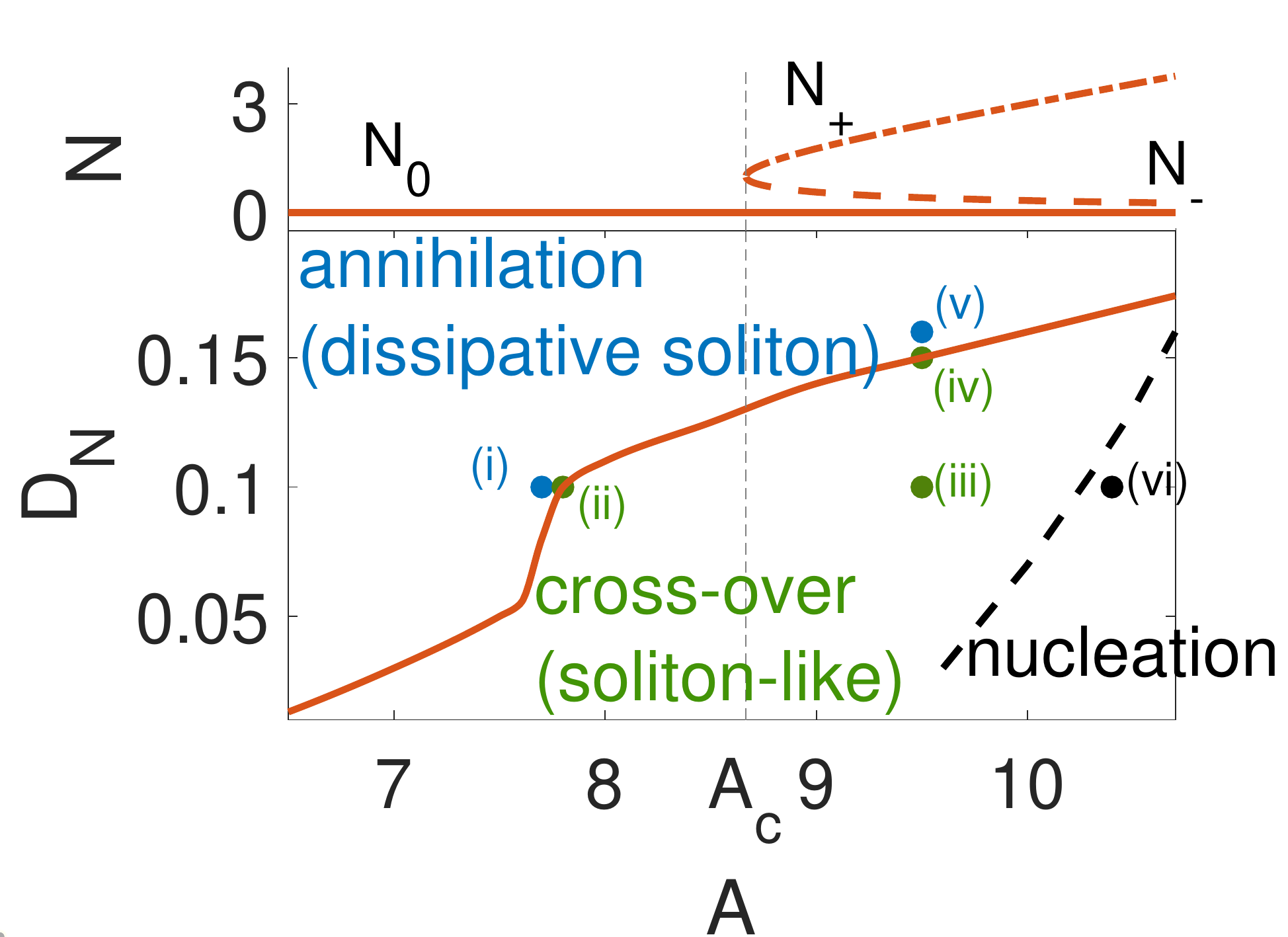} \quad \quad \quad
	(b)\includegraphics[width=0.4\textwidth]{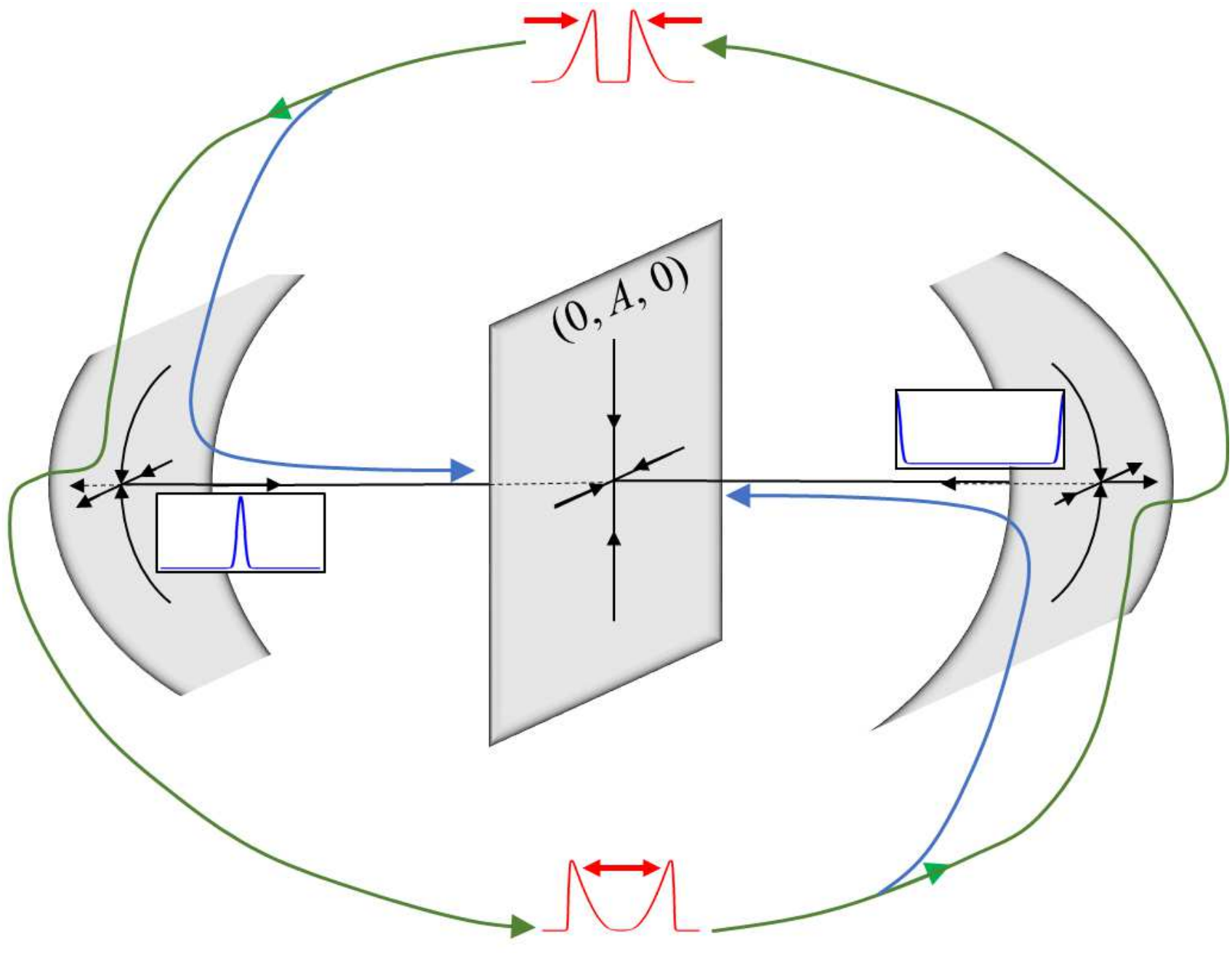}
	{\setlength{\unitlength}{0.1\textwidth}
		\begin{picture}(10,2)
		\put(0,0){\includegraphics[width=0.19\textwidth]{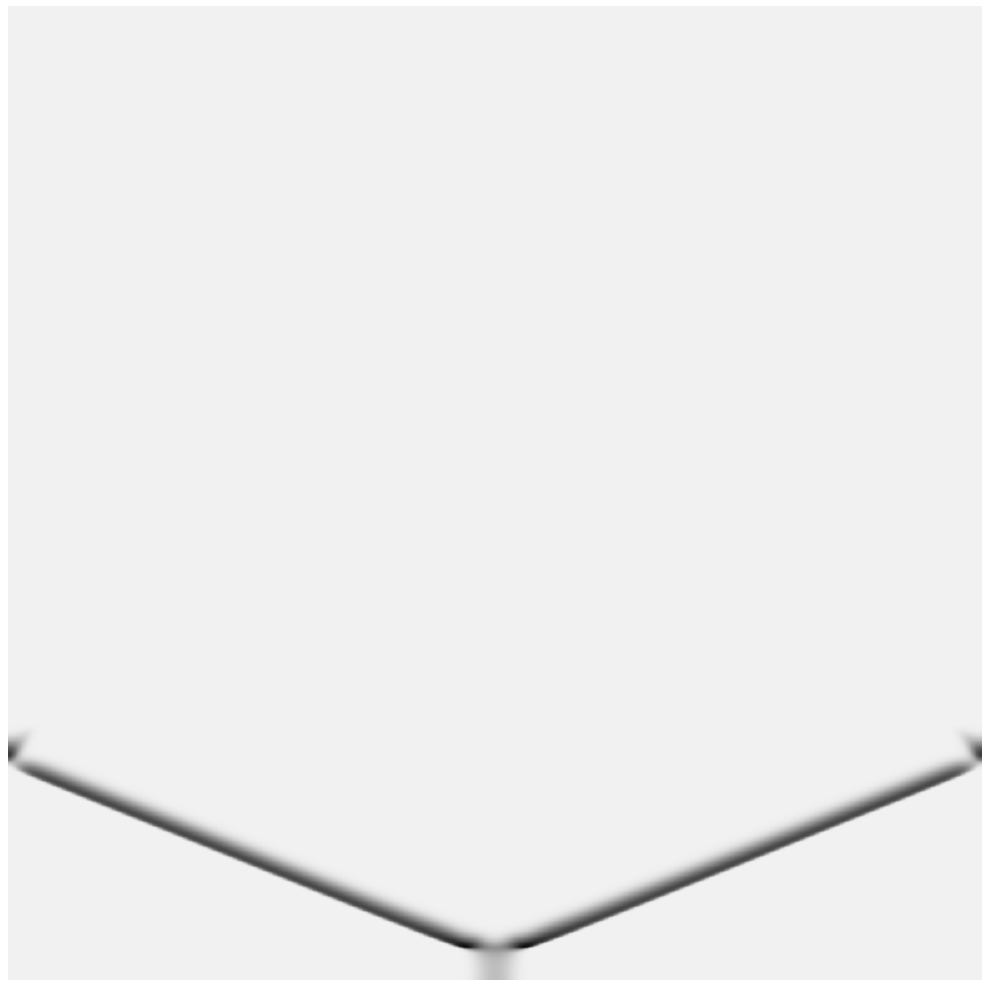}}
		\put(0.1,0.1){(i)}
		\put(0.1,1.5){$\uparrow t$}
		\put(2,0){\includegraphics[width=0.19\textwidth]{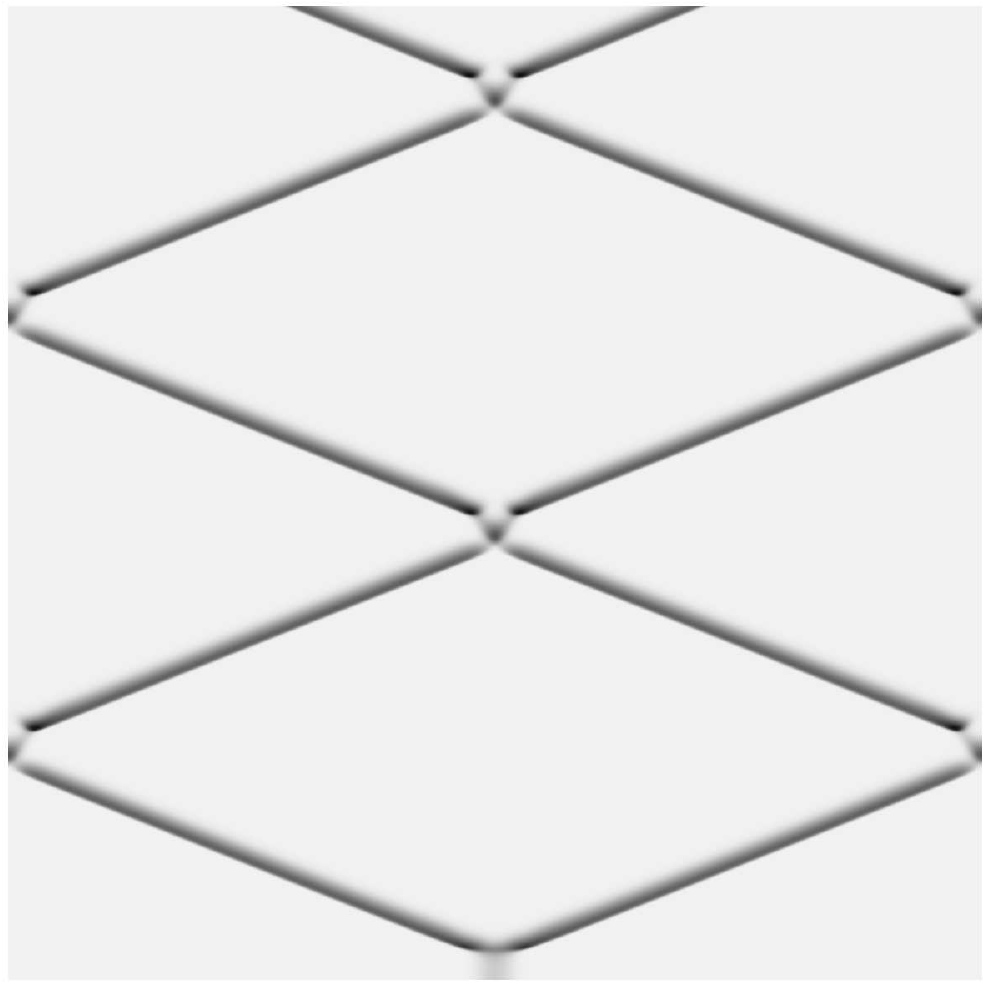}}
		\put(2.1,0.1){(ii)}
		\put(4,0){\includegraphics[width=0.19\textwidth]{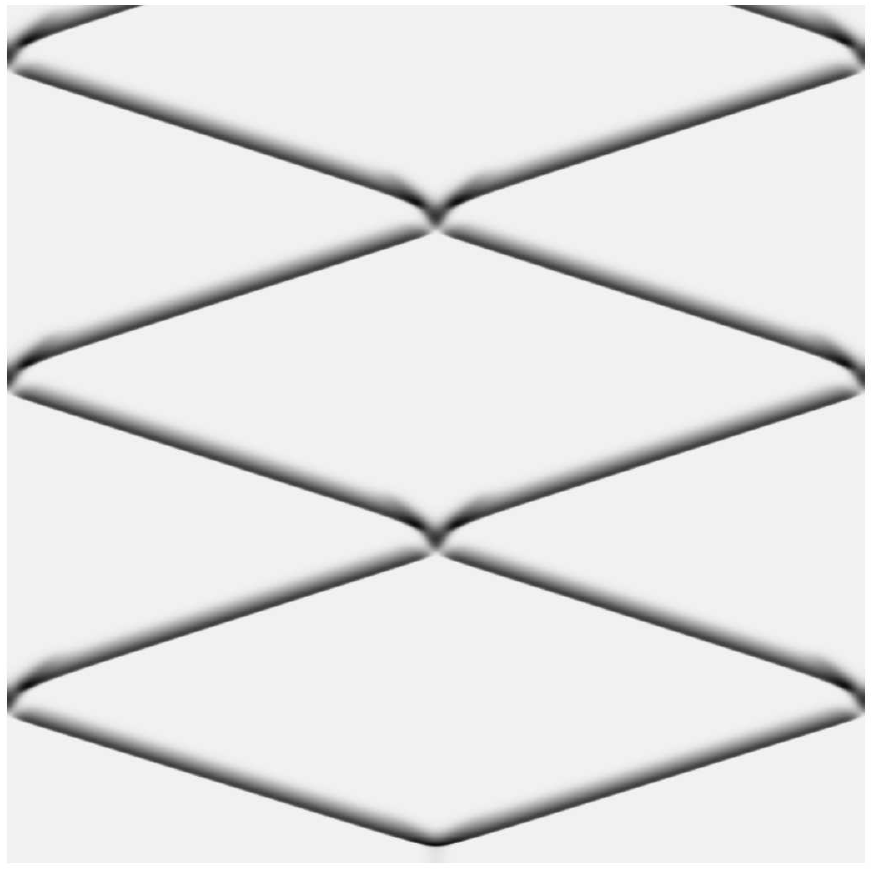}}
		\put(4.1,0.1){(iii)}
		\put(6,0){\includegraphics[width=0.19\textwidth]{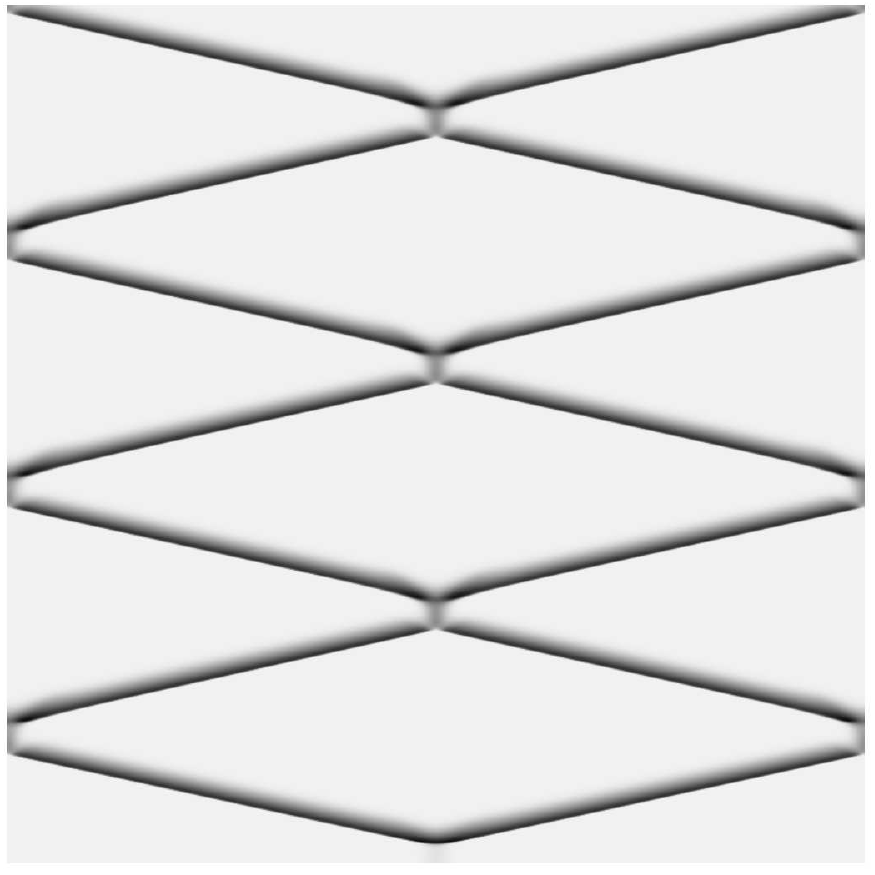}}
		\put(6.1,0.1){(iv)}
		\put(8,0){\includegraphics[width=0.19\textwidth]{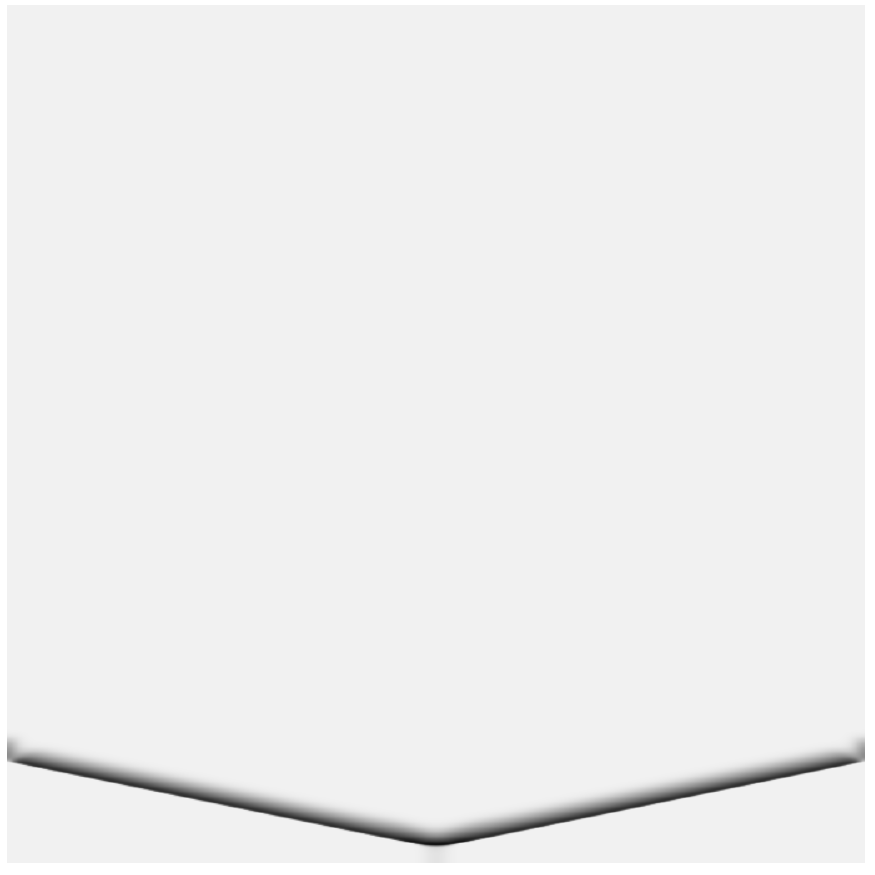}}
		\put(8.1,0.1){(v)}
		\end{picture}}
	{\setlength{\unitlength}{0.1\textwidth}
		\begin{picture}(10,2)
		\put(0,0){\includegraphics[width=0.19\textwidth]{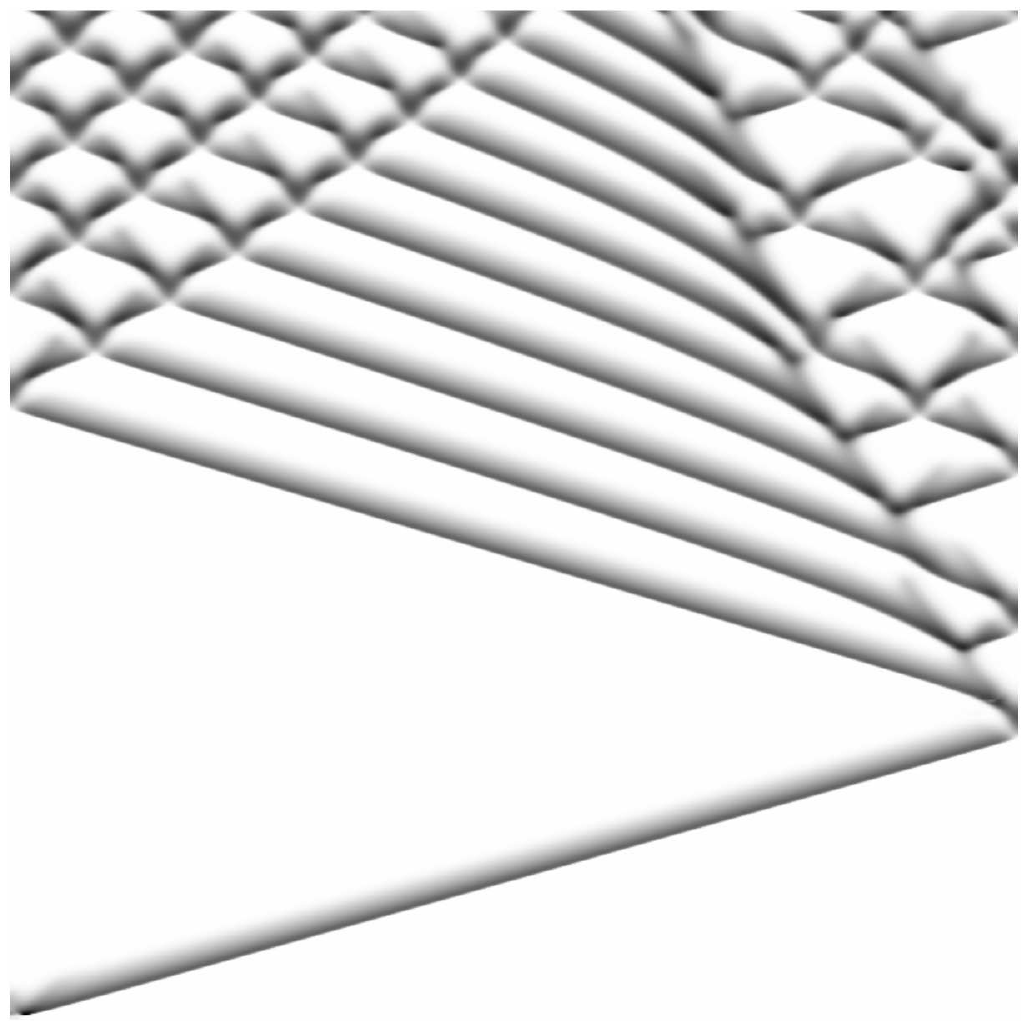}}
		\put(1.6,0.1){(vi)}
		\put(1.6,0.53){{\color{blue}$\boldsymbol{\Rightarrow$}}}
		\put(1.8,0){\includegraphics[width=0.2\textwidth]{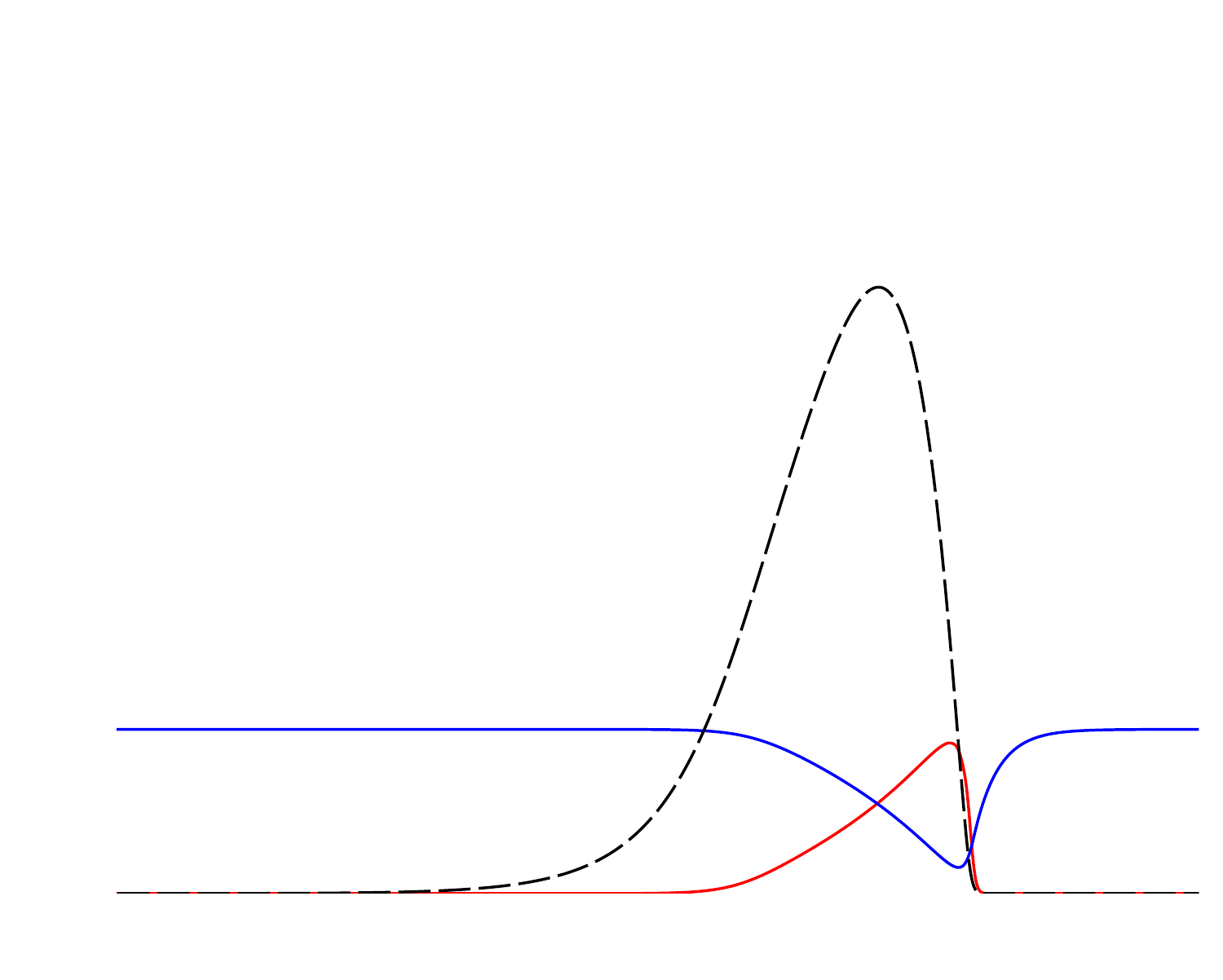}}
		\put(2.2,1.5){$t$=100}
		\put(3.3,1.0){{$\boldsymbol{\rightarrow$}}}
		\put(2.1,0.16){{\color{red}$N$}}
		\put(2.1,0.45){{\color{blue}$S$}}
		\put(2.9,0.6){{\color{black}$I$}}
		\put(3.8,0){\includegraphics[width=0.2\textwidth]{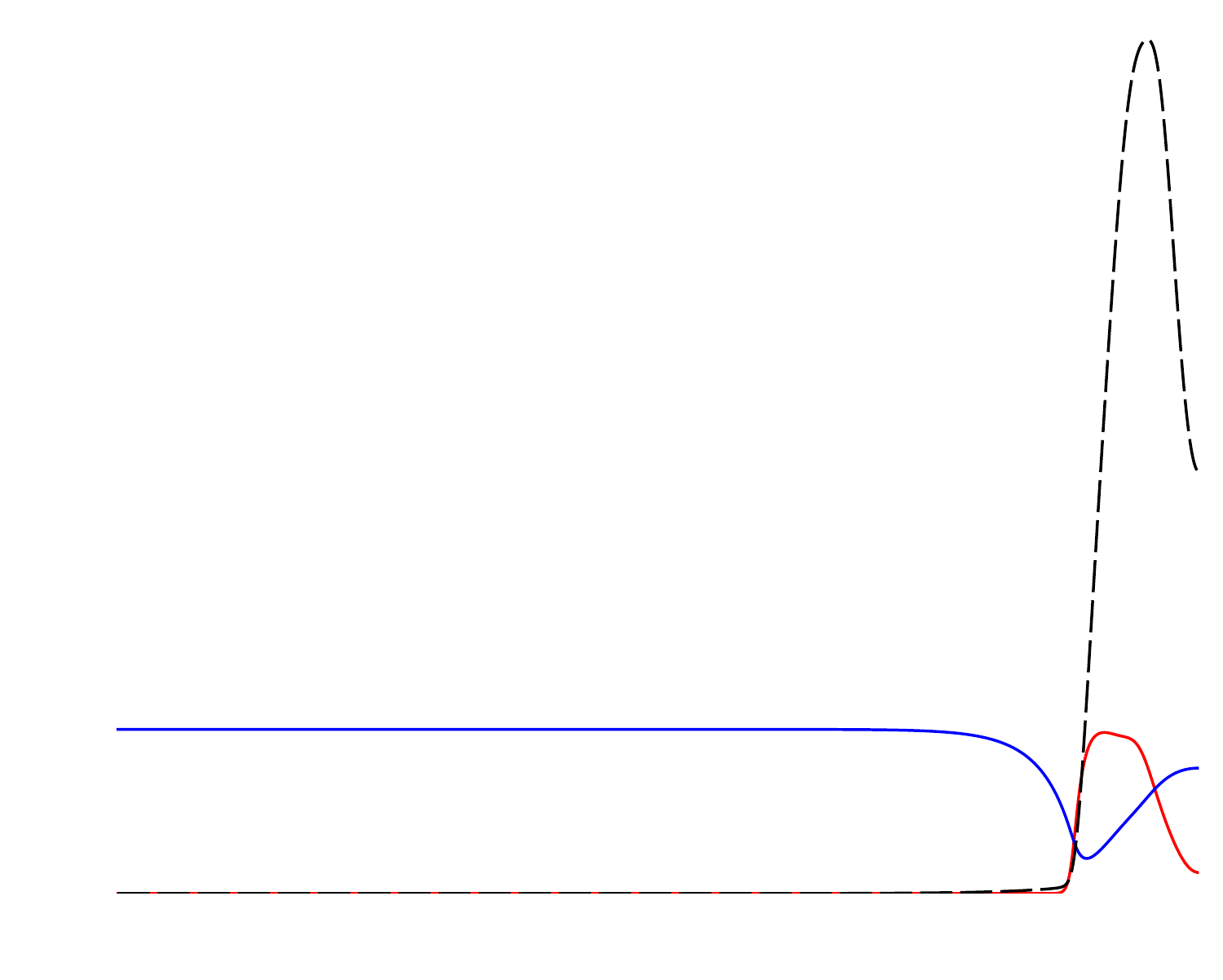}}
		\put(4.1,1.5){$t$=130}
		\put(5.4,1.0){{$\boldsymbol{\leftarrow$}}}
		\put(5.8,0){\includegraphics[width=0.2\textwidth]{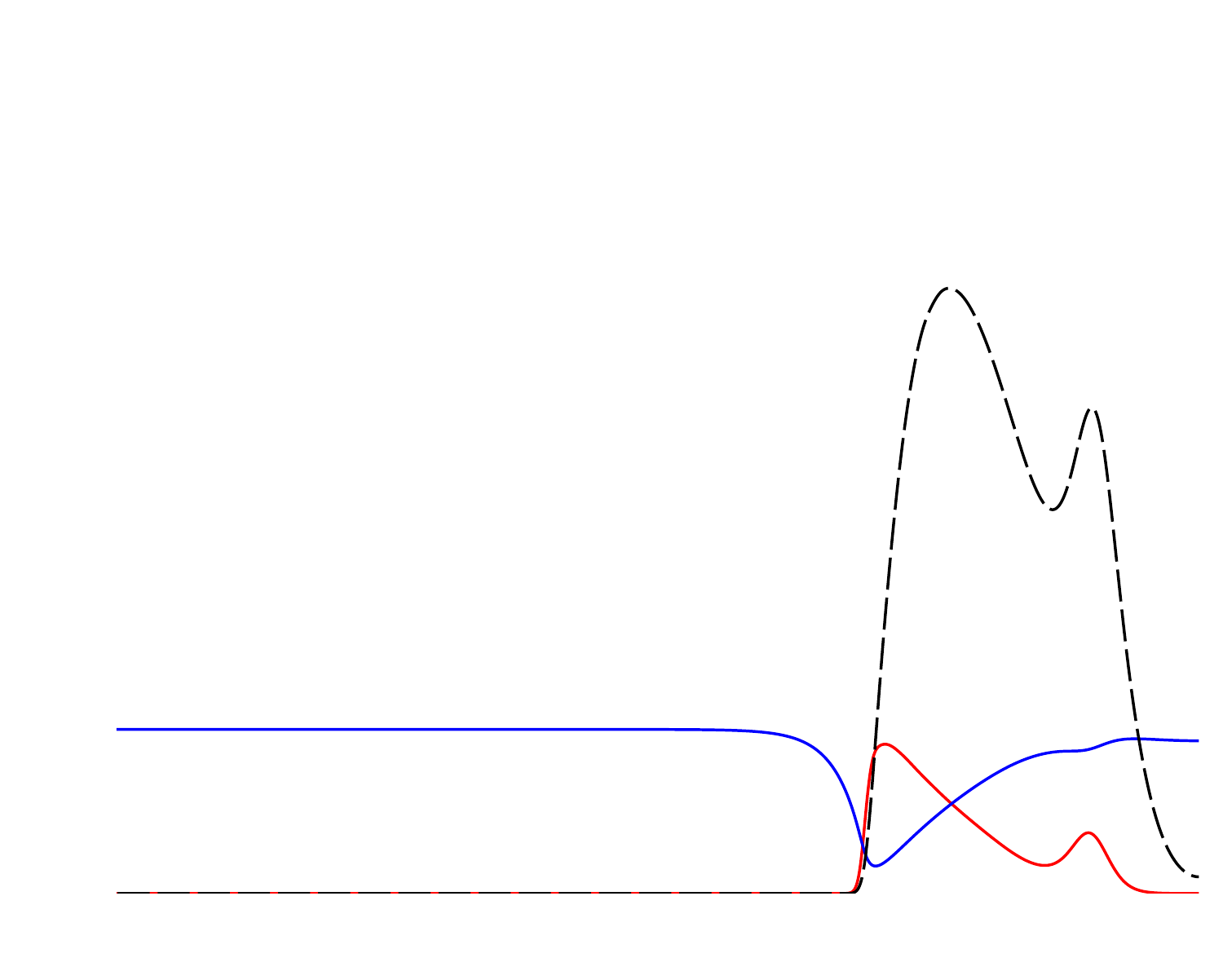}}
		\put(6.1,1.5){$t$=142}
		\put(7.08,1.0){{$\boldsymbol{\leftarrow$}}}
		\put(7.8,0){\includegraphics[width=0.2\textwidth]{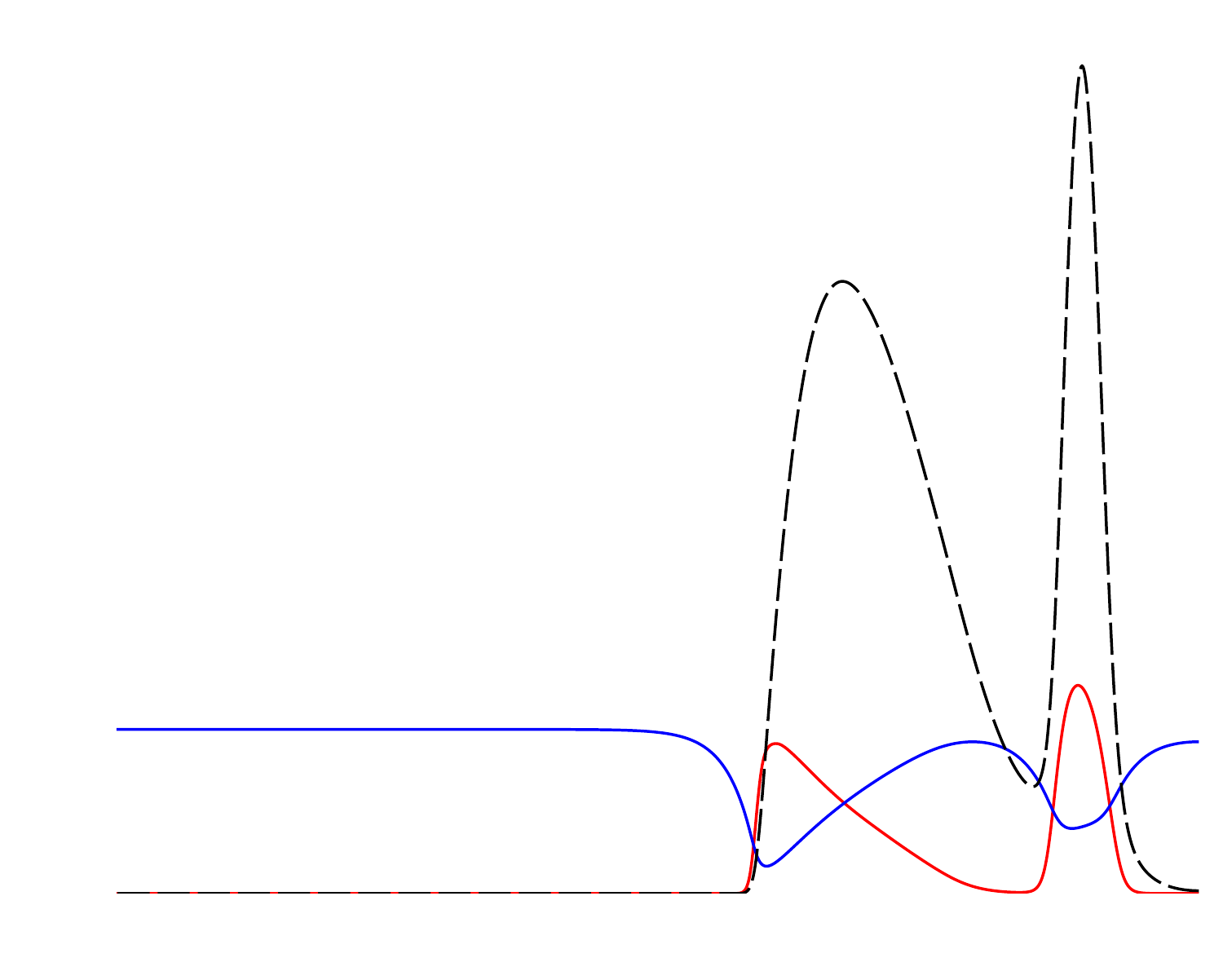}}
		\put(8.1,1.5){$t$=148}
		\put(8.9,1.0){{$\boldsymbol{\leftarrow$}}}
		\end{picture}}
	\caption{(Color online) (a) Bifurcation diagram for uniform states (top panel), where solid line indicates linear stability, dashed line indicates instability to uniform perturbations, whereas dashed-dotted line indicates instability to uniform Hopf and $A_c$ marks the location of the saddle node above which three solutions coexist. Parameter plane (bottom panel) reflecting distinct behaviors after collision of two excitable pulses and also the nucleation region (below dashed line). (b) Schematic representation of a simplified geometrical configuration for the collision process that is based on~\cite{argentina2000head}, see also text for details. The gray shaded regions represent the essential directions of manifolds, which are of much higher dimensional. The middle region is related to the fixed point $\mathbf{P}_0$ while the right-left domains to the nucleation droplet, as shown in the insets, respectively. The top-bottom insets, show  propagating pulses and soliton-like (outer path) or annihilation (inner path) behavior. (i)-(vi) Space--time plots computed by direct numerical integration of~\eqref{eq:RD}, with dark color indicating larger amplitude of $N$, domain size $x \in [0,100]$, time interval $t \in [0,400]$, and Neumann (no-flux) boundary conditions, where: (i) $A=7.7$, $D_{\text{N}}=0.1$, (ii) $A=7.8$, $D_{\text{N}}=0.1$, (iii) $A=9.5$, $D_{\text{N}}=0.1$, (iv) $A=9.5$, $D_{\text{N}}=0.15$, (v) $A=9.5$, $D_{\text{N}}=0.16$, (vi) $A=10.4$, $D_{\text{N}}=0.1$. In (i)-(v) the initial conditions are linear perturbations of an unstable steady--state solutions at respective values (e.g., Fig.~\ref{fig:droplet}). The spatial profiles right to (vi), show explicitly nucleation of new excitation after the first collision (see arrow in space-time plot) at indicated times, where dark arrows indicate directions of motion and $x\in[50,100]$. Other parameters: $D_{\text{I}}=0.001$, $k_N=2$, $k_N=0.3$, $A_c=8\,\sfrac{2}{3}$.}
	\label{fig:bif}
\end{figure*}
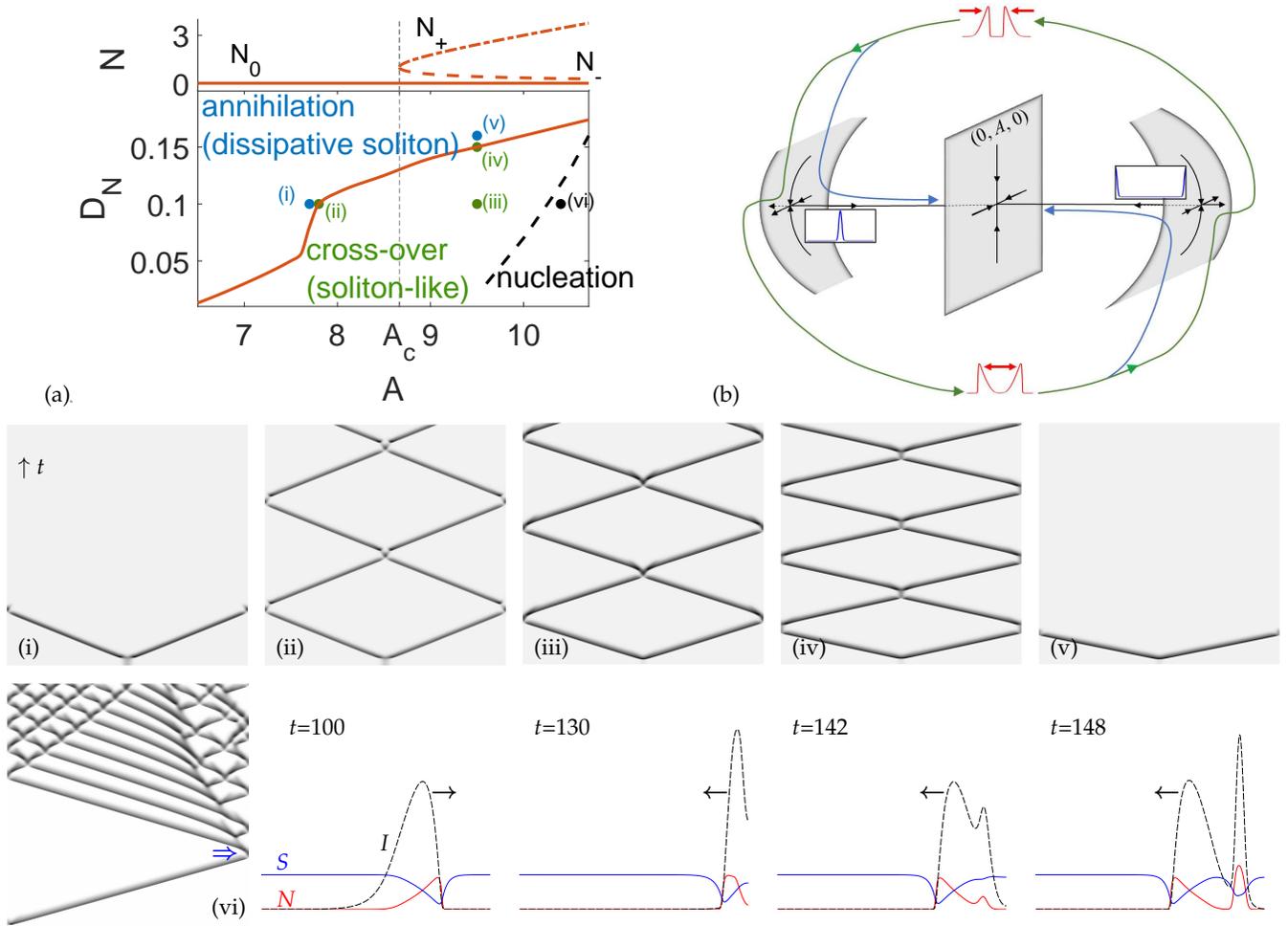

Our interest is in pulses, a situation that requires linear stability of a uniform solution. Eqs.~\ref{eq:RD} admit three uniform solutions $\mathbf{P} \equiv (N,S,I)^T$:
\[
\mathbf{P}_0=(0,A,0)^T,
\]
\[
\mathbf{P}_\pm=(N_\pm,A-N_\pm,\sfrac{k_\text{N}}{k_\text{I}} N_\pm)^T,
\]
where, $N_\pm=\dfrac{1}{2}\Bra{A-\sfrac{k_\text{N}}{k_\text{I}}\pm \sqrt{\bra{A-\sfrac{k_\text{N}}{k_\text{I}}}^2-4}}$
and superscript $T$ stands for transpose. Beyond the saddle--node (fold) bifurcation at
\[
A>A_c=\sfrac{k_\text{N}}{k_\text{I}}+2,
\]
see Fig.~\ref{fig:bif}{(a), top panel}, the solutions $\mathbf{P}_{\pm}$ appear: $\mathbf{P}_{-}$ is unstable by definition, while linear stability analysis of $\mathbf{P}_{+}$ to uniform perturbations shows that it is also unstable to Hopf oscillations, already from the saddle--node bifurcation point.

Next, we check linear stability of $\mathbf{P}_{0,+}$ to nonuniform perturbations on an infinite domain~\cite{Cross1993},
\begin{equation*}\label{eq:linear}
	\mathbf{P}-\mathbf{P}_{0,+} \propto e^{\sigma t+iqx} +\mathrm{c.c.},
\end{equation*}
where, $\sigma$ is the growth rate of perturbations that are characterized by wavenumbers $q$ and c.c. stands for complex conjugate. We find that solution $\mathbf{P}_0$ continues to be linearly stable and does not lie in a proximity to any linear oscillatory instability since all parameters are positive, with dispersion relations:
\begin{subequations}
	\begin{eqnarray}
	\sigma_N &=&-1-D_\text{N}q^2, \\
	\label{dispersion}
	\sigma_S &=&-q^2, \\ 
	\sigma_I &=&-k_\text{I}-D_\text{I}q^2.
	\end{eqnarray}
\end{subequations}
While the solution $\mathbf{P}_+$ was found to be unstable to uniform perturbations, we find that it is unstable also to traveling waves, i.e., non-vanishing imaginary part of its eigenvalue $\sigma_{+}$. However, these traveling waves are beyond the scope of our interest here and therefore, not shown.

Notably, the signature of mass-conservation is reflected in the persistence of the neutral mode $\sigma_S(q=0)=0$ (Eq.~\ref{dispersion}), which indicates a respective mass exchange between $N$ and $S$. This property is absent in the typical RD system without mass-conservation, e.g., FHN~\cite{argentina2000head}, and in what follows, we show that it plays an essential role during the collision of two counter propagating pulses, as shown in Fig.~\ref{fig:bif}.

\section{Spatial dynamics and the collision zone}	

As has been shown by Argentina {\it et al.}~\cite{argentina2000head}, information about the possible behavior after a collision between pulses can be deduced by looking at the geometric structure of the collision zone, i.e., by understanding the instability of coexisted symmetric steady--state solution to which the propagating pulses attempt to emerge at the collision, a.k.a \emph{nucleation droplet}. Such spatially localized states are associated with an intersection of two-dimensional stable and unstable manifolds in space~\cite{champneys1998homoclinic,knobloch2015spatial,knobloch2016localized}, meaning that pulse solutions connect asymptotically to $\mathbf{P}_0$ at $x \to \pm \infty$. {In what follows we find it useful to employ the nucleation droplet methodology, while noting that more advanced methods that include oscillations, bistability, and global connections, the so-called theory of scattors and separators, were developed by Nishiura \textit{et al.}~\cite{nishiura2003scattering,nishiura2003dynamic,teramoto2004phase}.} 

To identify the geometric structure of the nucleation droplet, we rewrite~\eqref{eq:RD} as a set of ordinary differential equations in a co-moving frame $\xi=x-c t$, where $c$ is the pulse propagation speed:
\begin{subequations}\label{eq:RDsp}
	\begin{align}
		\frac{\text{d} N}{\text{d} \xi}&=u,\\
		\frac{\text{d} S}{\text{d} \xi}&=v,\\
		\frac{\text{d} I}{\text{d} \xi}&=w,\\
		{D}_\text{N}\frac{\text{d} u}{\text{d} \xi}&={N-\frac{N^2 S}{1+I}}-c u,\\
		\frac{\text{d} v}{\text{d} \xi}&={\frac{N^2 S}{1+I} - N-c v},\\
		{D}_\text{I}\frac{\text{d} w}{\text{d} \xi}&={k_\text{I} I}-{k_\text{N} N}-c w,
	\end{align}
\end{subequations}
and perform linear (asymptotic) analysis in space~\cite{champneys1998homoclinic,knobloch2015spatial,knobloch2016localized}:
\begin{equation*}\label{eq:linear_sp}
	\mathbf{P}-\mathbf{P}_{0} \propto e^{\lambda \xi} +\mathrm{c.c.} \,.
\end{equation*}
The resulting spatial eigenvalues are:
\[
\lambda_0=0,
\]
\[
\lambda_{\text{c}}=-c,
\]
\[
\lambda_{\pm \text{N}}=\pm D^{-1}_{\text{N}},
\]
\[
\lambda_{\pm \text{I}}=\frac{-c\pm \sqrt{{c^2 +4D_{\text{I}} k_{\text{I}}}}}{2D_{\text{I}}}.
\]

\begin{figure}[tp]
	(a)\includegraphics[width=0.4\textwidth]{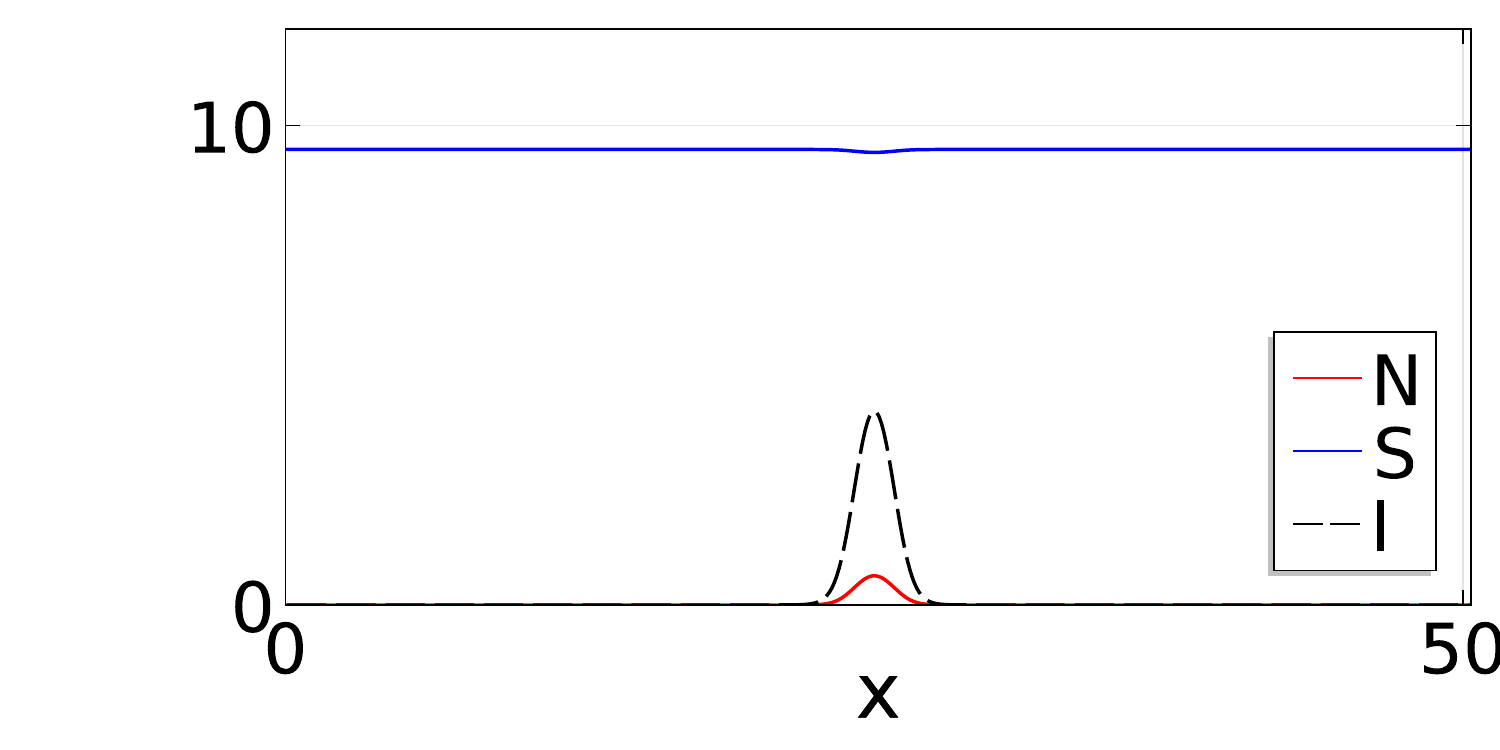}
	(b)\includegraphics[width=0.4\textwidth]{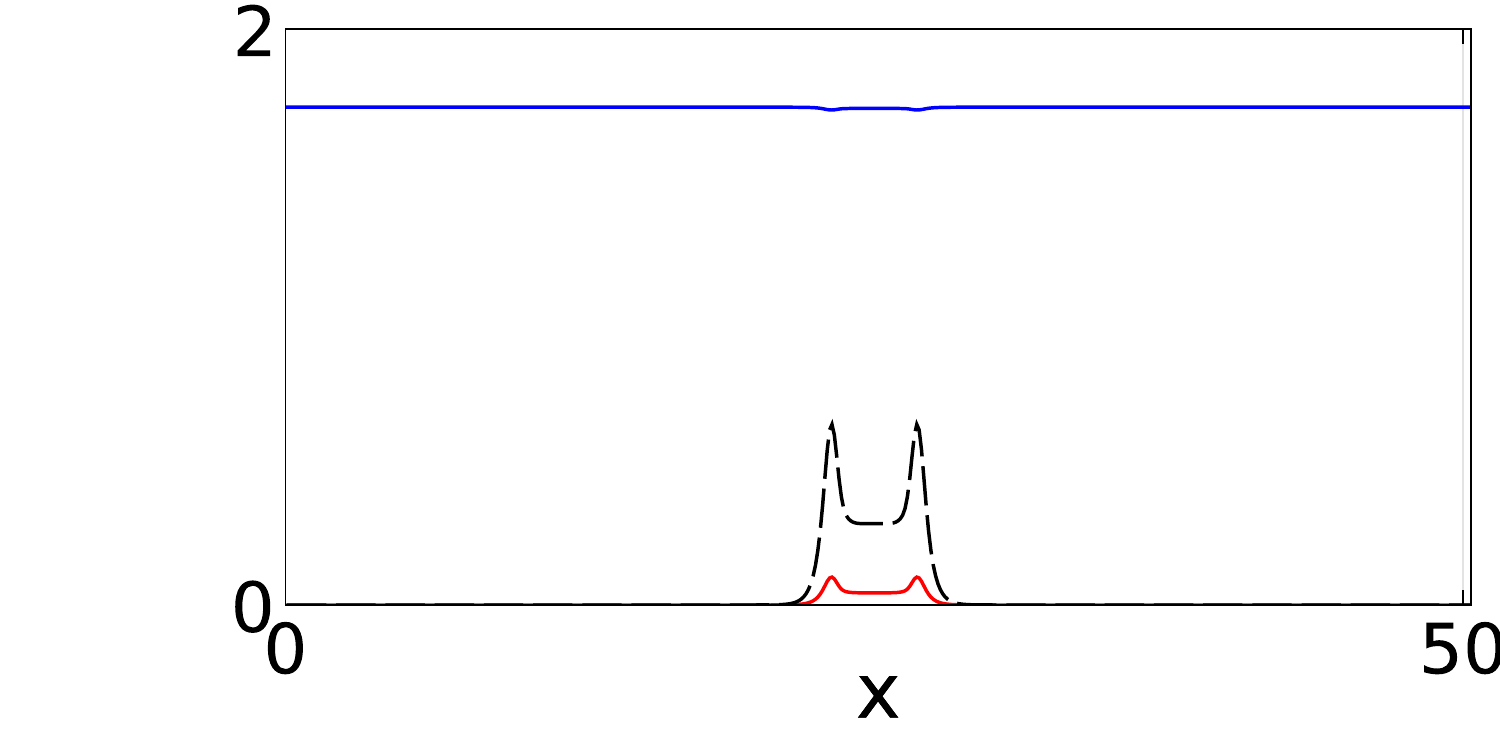}
	\caption{(Color online) (a) A typical unstable steady-state solution to~\eqref{eq:RD} computed numerically via boundary value problem at $A=9.5$, $D_{\text{N}}=0.1$, i.e., point (iii) in Fig.~\ref{fig:bif}. (b) Eigenfunction of steady--state solution (a) that corresponds to the neutral eigenvalue.
	}
	\label{fig:droplet}
\end{figure}

Inspection of the eigenvalues shows two distinct feature as compared to the FHN system: ({\it i}) The eigenvalues are all real so that the hyperbolic intersection at $\mathbf{P}_0$ results in monotonic tails of the pulses (Fig.~\ref{fig:bif}(vi), $t=100$), unlike in the FHN case, where the tails are oscillatory due to complex eigenvalues (which also indicates proximity to a Hopf onset in the FHN case), ({\it ii}) in addition to the 2D stable and unstable manifolds (as for FHN), an additional 2D manifold coexists, and specifically it becomes neutral at $c=0$, where $\lambda_0=\lambda_{\text{c}}=0$. The eigenvalue $\lambda_0=0$ is a signature of the mass conservation while $c=0$ implies a spatially symmetric (static) pulse solution. In fact, the 2D manifold that is associated with $\lambda_{\pm \text{I}}$ is not essential and all the results persist also for $D_{\text{I}}=0$, for which $\lambda_{\text{I}}=k_{\text{I}}/c$. {Indeed, a computation for $D_{\text{I}}=0$, shows that both the nucleation droplet and the eigenfunction are essentially identical, i.e., exhibit the same shape as those in Fig.~\ref{fig:droplet}. Moreover, elimination of $\lambda_{\pm \text{I}}$ indicates that the spatial picture is insensitive with respect to the rates in the $I$ field, leaving $D_N$ as the only control parameter. From a physical point of view $D_N$ controls the rate of transport of actin monomers so that the higher is the total number of actin monomers the larger must $D_N$ be to trigger annihilation of pulses (see Fig.\ref{fig:bif}a).} Hence, the intersection with 2D neutral manifold which arises from mass-conservation, i.e., the constraint that it imposes on the 2D manifold of $\lambda_{\pm\text{N}}$, adds distinct features as compared to non-conserved RD system.

\section{Numerical analysis and interpretation}
After identifying the necessary conditions for the nucleation droplet, we turn to numerical verification by solving Eqs.~\ref{eq:RDsp} as a standard boundary value problem with $c=0$. Indeed, we obtain a spatially symmetric stationary pulse solution that asymptotes to $\mathbf{P}_0$ as $x \to \pm \infty$, as shown in Fig.~\ref{fig:droplet}(a). Solving next the eigenvalue problem for the obtained pulse solution, we find that it is indeed linearly unstable. However, we also find that the critical information lies in the neutral eigenvalue, for which the associated eigenfunction is localized, as shown in Fig.~\ref{fig:droplet}(b). Notably, we get the same nucleation droplet and localized eigenfunction in the absence of diffusion of the inhibitor, for $D_\text{I}=0$. In the absence of mass conservation, this type of neutral eigenvalue and localized eigenfunction are absent. In the FHN model 
the nucleation droplet goes through an Andronov-Hopf instability~\cite{argentina2000head}, and is oscillatory.

The form of the localized eigenfunction that qualitatively persists over the entire parameter space (Fig.~\ref{fig:droplet}(b)), implies a splitting process of the nucleation droplet, and the initiation of soliton-like behavior over a wide range of parameters. To verify both nucleation and post-collision behavior, we use the nucleation droplet as an initial condition for direct numerical integration of Eqs.~\ref{eq:RD}. The parameters for the calculations are indicated by the points (i)-(vi) in Fig.~\ref{fig:bif}(a), while the space-time plots are shown with their respective numeral label. We chose examples from regimes where two colliding pulses (i,v) annihilate (dissipative excitable solitons) or (ii-iv) persist (soliton-like) upon collision.

However, mass-conservation apparently, holds another post-collision feature, which is related to the spontaneous emergence of a new symmetric pulse at the tail of each pulse after the collision takes place. The behavior is marked as \emph{nucleation} region in Fig.~\ref{fig:bif}(a). Panel (vi) shows a space-time plot at which these nucleation processes are formed giving rise to a persistent wavy pattern. The profiles on the left of panel (vi) show a single boundary-collision event (see double arrow in (vi)) after which a new symmetric pulse emerges. Formation of this new pulse is related to mass-conservation. {After the collision, the lagging inhibitor (after the leading pulse front, $N$) is concentrated in space to high values ($t=130$). This over-shoot in $I$ creates a steep decrease also at the back of the pulse, much faster than the steady-state exponential decay (see $t=100$, before the reflection). The depolymerized mass of $N$ is conserved and converted into a high density of monomers $S$, which provide the substrate for the nucleation of a new (almost symmetric) pulse ($t=142$). This nucleation can then either decay (the soliton-like behavior as in (ii)-(iv)) or grow ((vi), $t=148$), and then split into two new counter-propagating pulses, which subsequently generate the pattern shown in (vi). Naturally, a similar nucleation mechanism does not occur in RD media without mass conservation since the deformation of the pulses is not constrained, and the nucleation droplet is oscillatory, which can at most give rise to multiple oscillating waves upon pulse collisions~\cite{argentina2000head} or chaotic dynamics~\cite{nishiura2003dynamic}.}

To summarize the analysis, we follow for convenience the schematic (and in our case also over simplified) geometrical representation by Argentina \textit{et al.}~\cite{argentina2000head}. In Figure~\ref{fig:bif}(b), we show the essential manifolds for pulses that collide either at the middle of the domain or at the boundaries, where the main difference as compared to RD media  without mass-conservation~\cite{argentina2000head}, is the center manifold for the nucleation droplet, see the most left and right profiles.

\section{Discussion}
{Mass-conservation constraints are particularly important in enclosed systems, such as biological cells, for phenomena that occur on time-scales which are fast compared to the rate at which global protein content changes through protein biosynthesis and degradation. Indeed, the importance of mass-conservation for cellular phenomena has recently attracted increasing attention in the context of reaction-diffusion modeling of for example, intracellular patterns~\cite{miura2007cortical,holmes2012regimes,yochelis2015self,yochelis2016reaction,halatek2018rethinking,fai2019length}.}
We have shown in this study that RD media with {mass} conservation can support rich {spatiotemporal dynamics} following pulse collisions: annihilation, crossover, and ``birth'' of new pulses after crossover. {Due to mass-conservation, this behavior is robustly observed over a wide range of parameters, and solely triggered by mass transport, which is determined by the diffusion coefficient parameter ($D_N$) in our model equations.}
No special conditioning, such as proximity to a bifurcation point, non-locality, or cross--diffusion, is required, in contrast to RD-type models, which do not employ explicit mass-conservation~\cite{whitelam2009transformation,dreher2014spiral,miao2019wave,alonso2018modeling}. This implies that collisions can, in fact, be viewed as organizing centers of coexisting distinct outcomes. For example, it has been shown that in excitable media, \emph{single} focal perturbations may generate distinct multiple-integer pulse trains, due to proximity to the Hopf-Shil'nikov bifurcation that allows coexistence of both traveling and standing waves, along with excitable pulses~\cite{YKXQG:08,yochelis2015origin}.

These phenomena are specifically relevant to actin waves that occur in a wide range of cell types~\cite{beta_intracellular_2017,inagaki2017actin}. Although still under debate, their role has been associated with essential cellular functions, such as polarity formation, motility, and phagocytosis. Sustained wave activity may thus become a key requirement for proper cell function, and is even associated with cancerous phenotypes~\cite{Itoh2012,Hoon2012}. Soliton-like collisions and pulse nucleation that robustly emerge in a mass-conserved system can be seen as a strategy to maintain prolonged wave activity without depending on local heterogeneities or actively introduced nucleation events. In this regime, waves persist and replicate, in contrast to a ``classical'' excitable media, where pulses mutually annihilate upon collision or decay at the boundaries. Moreover, we may also envision that cells control their level of intracellular wave activity by gradually shifting between regimes of pulse annihilation and soliton-like behavior.

We therefore exemplified our findings using a model of intracellular actin polymerization that describes the dynamics of circular wave patterns at the dorsal membrane of adherent cells~\cite{bernitt2017fronts}.
We believe that the effects of mass conservation on pulse collision dynamics presented here will stimulate further progress in the modeling of actin waves and will thus advance our understanding of intracellular wave patterns in general. Moreover, they will also impact studies of non-biological media, such as catalytic surface reactions and electrochemical systems that exhibit solitary waves~\cite{rotermund1991solitons,von1998subsurface,krischer2001fronts}, where surface coverages often obey similar conservation characteristics~\cite{avital2018two}, and polymerization in active gels~\cite{levernier2019spontaneous}.

%

\end{document}